# To Study the Effect of Grating Length on Propagating Modes in Bragg Filters with $Al_XGa_{1-X}N/GaN$ Material Composition

Sourangsu Banerji
*Department of Electronics & Communication Engineering,
RCC Institute of Information Technology, India
sourangsu.banerji@gmail.com*

*Abstract*

*In this paper, the forward and backward propagating modes in an optical waveguide structure namely the fiber Bragg filter also considered as a one dimensional photonic crystal, is analytically computed as a function of grating length for coupled optical modes. $Al_xGa_{1-x}N/GaN$ material composition is considered as a unit block of the periodic organization, and refractive index of $Al_xGa_{1-x}N/GaN$ is taken to be dependent on material composition, bandgap and operating wavelength following Adachi's model. Expressions for propagating wave have been derived using coupled mode theory. Simulated results help us to study the propagation of forward and backward wave propagating modes inside fiber and waveguide devices.*

## 1. Introduction

Photonic crystal is a periodic multilayer arrangement of dielectric materials having alternating higher and lower dielectric constants; where propagation of electromagnetic wave is restricted for a few particular frequency bands and allowed in other regions. It is considered a revolution from the communication engineering stand-point due to its ability of displaying arbitrarily different dispersions for the propagation of electromagnetic waves. This is possible by virtue of the concept of photonic bandgap, and researchers already studied 1D, 2D and 3D structures for different applications [1]. Photonic crystals have been already used to construct optical waveguides [2], photonic band-edge laser [3], high efficient LED [4], filter [5], switches [6]. Photonic crystal fibres are developed on the physics required for the primary purpose of optical communication [7][20], optical nonlinearity [8], integrated photonics [9], sensing [10], high power technology [11], and quantum information science [12]. This novel microstructure already replaced conventional optical fiber for efficient communication.

In Section 2 of this paper, we seek to develop a model for a device that is of importance for microphotonics. The "device" we consider is simply the optics required for coupling of light into, and out of the optical fibers. We discuss the concept of mode matching and quantify the coupling losses. We then develop a perturbation theory, called coupled-mode theory, useful for modeling of optical devices based on the coupling and interference between two or more propagating modes. We will use coupled mode theory to describe Bragg reflectors. These devices can be implemented both as fiber components as well as multi-layer film stacks, and play important roles as filters and reflectors in many optical systems. Bragg reflectors can also be thought of as one-dimensional Photonic Crystals. Now, among the different photonic bandgap structures, 1D periodic photonic bandgap microstructures have been studied by various researchers in the last decade due to the advantage of theoretically analyzing optical characteristics near accurately with lack of confinement in two spatial dimensions.

Foteinopoulou analyzed the effect of surface defect on backward wave, and it is shown that surface mode manipulation is possible with dispersion [13]. Propagating wave analysis is useful for designing four-wave mixing analysis in nonlinear photonic crystal [14]. Suitable dielectric material is used to characterize modal dispersion in 1D crystal [15]. Computation of wave profile is much easier in this structure, and incorporation of semiconductor nanostructure makes it more interesting when filter characteristics is considered [16] including the effect of polarization of incident light.

Use of $Al_xGa_{1-x}N$/GaN material composition was taken up because as shown in [18] that the effect of carrier localization in undoped AlGaN alloys enhances with the increase in Al contents and is related to the insulating nature of AlGaN of high Al contents. The use of high Al-content AlGaN layer is also expected to increase the overall figure of merit of the AlGaN/GaN due to the combined advantages of enhanced band offset, lattice mismatch induced piezoelectric effect. Thus improving the material quality of high Al content AlGaN alloys is also of crucial importance for fabricating high performance AlGaN/GaN structures. The advantages of GaN can be summarized as ruggedness, power handling and low loss [17], [19].

Finally, in section 3 we study the forward and backward propagating wave modes in the waveguide structure which are analytically computed as a function of grating length for $Al_xGa_{1-x}N$/GaN composition with different coupling coefficients. Variation of wave profiles can provide the idea propagating modes as well as the optical characteristics of a fiber Bragg filters. Lastly we conclude our paper.

## 2. Mathematical Modeling

### 2.1. Coupling to Waveguides

Optical waveguides, in order to be used we somehow need to couple light in and out of them. There are several ways to excite the propagating modes on optical waveguides, but the simplest way is end-fire, in which we direct an optical field at the end of an optical fiber or waveguide. To maximize the coupling, we must maximize the overlap of the exciting field and the mode we want to excite as illustrated in Fig. 1

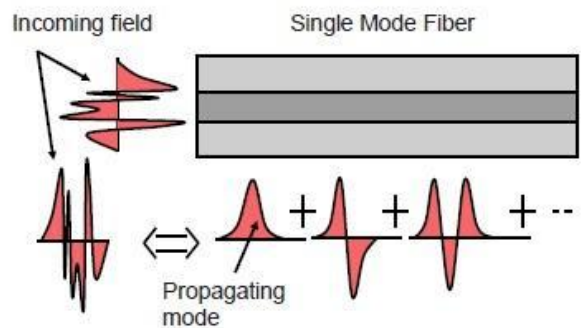

**Figure 1** End-fire coupling of an optical field to a waveguide. The profile of the incoming field should match the profile of the waveguide mode we wish to excite in order to maximize the coupling.

Assume that the incoming electric field can be expressed in terms of the guided modes on the fiber as follows:

$$\vec{E} = \sum_{guided} C_n \vec{E_n} + \int_{radiation} C_\beta \vec{E_\beta}\, d\beta \tag{1}$$

Using the Poynting theorem to express the propagating power in mode n, we find the following expression for the expansion coefficients

$$\int_{cross} (\vec{E_{in}} \times \vec{H_n^*}).dA = \int_{cross} [[\sum_{guided} C_n \vec{E_n} + \int_{radiation} C_\beta \vec{E_\beta}\, d\beta] \times \vec{H_n^*}].dA$$

$$= C_n \cdot \int_{cross} (\vec{E_n} \times \vec{H_n^*}).dA \Rightarrow C_n = \frac{\int_{cross} (\vec{E_{in}} \times \vec{H_n^*}).dA}{\int_{cross} (\vec{E_n} \times \vec{H_n^*}).dA} \tag{2}$$

Here orthogonality property of guided modes of an optical fiber has been used. The expression for the expansion coefficients of the incoming field when normalized are:

$$t_n = \frac{C_n}{\sqrt{\int_{cross} (\vec{E_{in}} \times \vec{H_{in}^*}).dA}} \sqrt{\int_{cross} (\vec{E_n} \times \vec{H_n^*}).dA}$$

$$= \frac{\int_{cross} (\vec{E_{in}} \times \vec{H_n^*}).dA}{\sqrt{\int_{cross} (\vec{E_{in}} \times \vec{H_{in}^*}).dA}} \cdot \frac{1}{\sqrt{\int_{cross} (\vec{E_n} \times \vec{H_n^*}).dA}} \tag{3}$$

The reflections at the waveguide interface complicate matters considerably because the reflection coefficients will depend on the propagation constants of the modes on the two sides of the interface. Finding exact solutions to this problem require that the incoming, as well as the transmitted fields, must be expressed in terms of modes with well-defined propagation constants. In most cases of practical interest, we may simply set the interface reflection and transmission coefficients to

$$r_{interface} = \frac{\beta_{in} - \beta_t}{\beta_{in} + \beta_t} = \frac{n_{eff,in} - n_{eff,t}}{n_{eff,in} + n_{eff,t}} \tag{4}$$

$$t_{interface} = \frac{2\beta_{in}}{\beta_{in} + \beta_t} = \frac{2n_{eff,in}}{n_{eff,in} + n_{eff,t}} \tag{5}$$

The total coupling coefficient from the incoming field into mode n, is then

$$t_n = \frac{2\beta_{in}}{\beta_{in} + \beta_t} \frac{\int_{cross} (\vec{E_{in}} \times \vec{H_n^*}).dA}{\sqrt{\int_{cross} (\vec{E_{in}} \times \vec{H_{in}^*}).dA} \sqrt{\int_{cross} (\vec{E_n} \times \vec{H_n^*}).dA}} \tag{6}$$

$$T_n = t_n t_n^* = [\frac{2\beta_{in}}{\beta_{in} + \beta_t}]^2 \frac{[\int_{cross} (\vec{E_{in}} \times \vec{H_n^*}).dA]^2}{\int_{cross} (\vec{E_{in}} \times \vec{H_{in}^*}).dA \int_{cross} (\vec{E_n} \times \vec{H_n^*}).dA} \tag{7}$$

These coupling formulas are significantly simplified if only the transversal components of the fields are considered. Most guided optical waves are close to Transversal Electro Magnetic (TEM), so ignoring the relatively small longitudinal part of the guided modes lead to insignificant over-estimation of the coupling coefficient in most, if not all, practical situations.

We observe that for TEM waves

$$H_x = \frac{-j}{\omega\mu_0}\frac{\partial E_y}{\partial z} = \frac{\beta}{\omega\mu_0}E_y \quad (8)$$

The coupling coefficient then becomes

$$t_n = \frac{2\sqrt{\beta_{in}}\sqrt{\beta_t}}{\beta_{in}+\beta_t}\frac{\int_{cross}(\vec{E_{in}}\ \vec{E_n^*}).dA}{\sqrt{\int_{cross}(\vec{E_{in}}\ \vec{E_{in}^*}).dA}\sqrt{\int_{cross}(\vec{E_n}\ \vec{E_n^*}).dA}} \quad (9)$$

The coupling coefficient then becomes

$$T_n = t_n t_n^* = [\frac{2\sqrt{\beta_{in}}\sqrt{\beta_t}}{\beta_{in}+\beta_t}]^2 \frac{[\int_{cross}(\vec{E_{in}}\ \vec{E_n^*}).dA]^2}{\int_{cross}(\vec{E_{in}}\ \vec{E_{in}^*}).dA\ \int_{cross}(\vec{E_n}\ \vec{E_n^*}).dA} \quad (10)$$

### 2.2. Coupling in Gratings

Gratings can also be used to facilitate phase matching between an incident optical field and a guided mode. To understand why, recall that a grating that is periodic in the z-direction, adds a propagation vector

$$K_s = q\frac{2\pi}{\Lambda} \quad (11)$$

to the optical field. In this formula q can be any integer (positive, negative, or zero). The additional k- vector provided by the grating can be designed to phase match a plane wave in the cladding of a waveguide to a guided mode. This is illustrated in Fig. 2. We see that in addition to the diffracted beam that is phase matched to the waveguide mode; there also exist other diffracted modes. Any power that is coupled into these modes is wasted i.e. it is not transferred into the guided mode. In designing couplers we therefore must pay careful attention to all diffraction modes to ensure effective coupling.

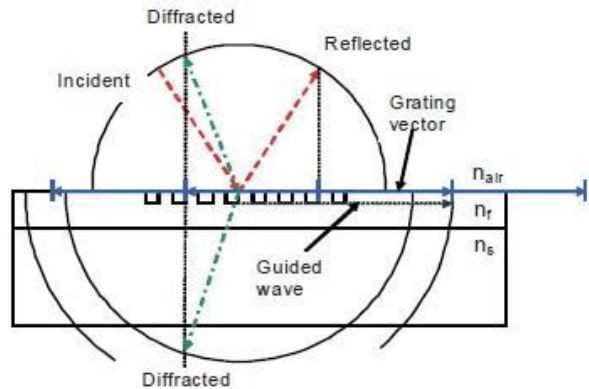

**Figure 2** Grating coupling. The incident light (dashed arrow) is coupled to several diffracted orders by the k-vectors of the grating (solid arrows). In addition to the reflected wave (dashed), and the guided wave (dotted), we also have one diffracted wave propagating (dot-dashed), and one in the substrate (dot-dashed).

### 2.3. Coupled Optical Modes

Propagating modes on optical waveguides can interact, and therefore be coupled, just as the mechanical and other types of oscillators. Consider the waveguide structure of Fig. 3. The individual waveguides of this structure are slab waveguides (i.e. they are of infinite extent in the y coordinate that is perpendicular to the plane). Their fields can therefore be expressed in rectangular coordinates without coupling between the fields along the coordinate axes. This greatly simplifies the problem, because it allows us to find each of the field components as a solution to the scalar wave equation. The full mode structure in the region where the waveguides are in close proximity is considerably more complex, so we are compelled to search for approximations that will give us analytical solutions.

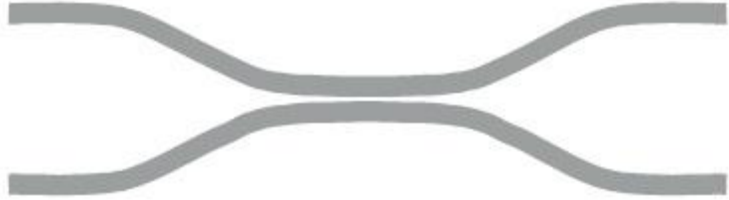

**Figure 3** Waveguide structure consisting of two waveguides that are brought into close proximity over a finite distance. In the area where the two waveguides are close, the modes of the waveguides will interact, which means that their propagation vectors will be shifted, and their modes modified.

Our starting point is the scalar wave equation:

$$\nabla^2 E_y(x,z) = \varepsilon\mu \frac{\partial^2 E_y(x,z)}{\partial t} \tag{12}$$

that is valid in all the regions of the waveguide. The electric permittivity, or dielectric constant, and the magnetic permeability are both time invariant, so the solutions we are interested in are independent of z, i.e. they are of the form

$$E_{yi}(x,z) = \frac{1}{2} A_i . u_{yi}(x) e^{-j(\beta_i z - \omega t)z} + c.c \tag{13}$$

If we now chose to first consider the fundamental TE mode of a symmetric waveguide, then the mode profile is given by

$$u_y(x,z) = \begin{cases} C \exp[-\gamma_c x] & x > 0 \\ C\left[\cos(\kappa_k x) - \frac{\gamma_c}{\kappa_k}\sin(\kappa_k x)\right] & -h < x < 0 \\ C\left[\cos(\kappa_k x) + \frac{\gamma_c}{\kappa_k}\sin(\kappa_k x)\right] \exp[-\gamma_s(x+h)] & x < -h \end{cases} \tag{14}$$

where is C is determined by the requirement

$$\int_S u_{yn}(x) u_{ym}(x)\, dx = \frac{2\omega\mu_0}{\beta_n}\delta_{mn} \tag{15}$$

As mentioned above, solving the wave equation on coupled waveguide structures is prohibitively hard and cannot be done analytically. Our approach to this problem is to consider the simpler structure, for which we can find the modes, and introduce the coupling as a perturbation of the polarization of the medium. To do that, we write constitutive relation for electric field as

$$\vec{D} = \varepsilon\vec{E} = \varepsilon_0\vec{E} + \vec{P} = \varepsilon\vec{E} + \vec{P}_{pert} \tag{16}$$

which means we can write the wave equation in the following way,

$$\nabla^2 E_y(x,z,t) = \varepsilon\mu \frac{\partial^2 E_y(x,z,t)}{\partial t} = \varepsilon\mu \frac{\partial^2 E_y(x,z)}{\partial t} + \mu \frac{\partial^2 P_{pert}(x,z,t)}{\partial t} \qquad (17)$$

First we set the perturbation to zero and find the modes of the unperturbed or uncoupled waveguide. The fields of the perturbed guide can be expressed in terms of these unperturbed modes

$$E_y(x) = \frac{1}{2}\sum_i A_i^+ u_i(x).\exp[-j(\beta_i.z - \omega.t)] + c.c$$
$$+ \frac{1}{2}\sum_i A_i^- u_i(x).\exp[-j(\beta_i.z + \omega.t)] + c.c$$
$$+ \frac{1}{2}\int_{k_0 n_s}^{k_0 n_f} A(\beta,z).u_\beta(x).\exp[-j(\beta.z + \omega.t)]d\beta + c.c \qquad (18)$$

where we have included both forward and backward traveling waves. If we carry the full field expansions, we haven't made any approximations, but the problem isn't simplified either. To make the problem feasible, we will assume that coupling to the radiation modes is negligible and therefore drop these modes from the expansion. Coupled-mode theory relies on this approximation, so negligible coupling to radiation modes can therefore be used as a criterion for when to apply coupled mode theory.

Now substitute the expanded solution back into the wave equation

$$\nabla^2\{\sum_i A_i^+ u_i(x).\exp[-j(\beta_i.z - \omega.t)]$$
$$+ \frac{1}{2}\sum_i A_i^- u_i(x).\exp[j(\beta_i.z + \omega.t)] + c.c\}$$
$$= \mu\frac{\partial^2}{\partial t}\{\sum_i A_i^+ u_i(x).\exp[-j(\beta_i.z - \omega.t)]$$
$$+ \frac{1}{2}\sum_i A_i^- u_i(x).\exp[j(\beta_i.z + \omega.t)] + c.c\} + \mu\frac{\partial^2 P_{pert}(x,z,t)}{\partial t}$$

$$\Rightarrow \sum_i [A_i^+\{-\beta_i^2 u_i(x) + \frac{\partial^2 u_i(x)}{\partial t} + \omega^2.\mu.\varepsilon.u_i(x)\}\exp[-j\beta_i z]$$
$$+ \{-2j\beta_i \frac{dA_i^+}{dz} + \frac{d^2 A_i^+}{dz^2}\}u_i(x)\exp[-j\beta_i z] + c.c] + \sum_i [A_i^-\{-\beta_i^2 u_i(x) + \frac{\partial^2 u_i(x)}{\partial t}$$
$$+ \omega^2.\mu.\varepsilon.u_i(x)\}\exp[-j\beta_i z] + \{-2j\beta_i \frac{dA_i^-}{dz} + \frac{d^2 A_i^-}{dz^2}\}u_i(x)\exp[j\beta_i z] + c.c]$$
$$= 2\exp[-j\omega t].\mu.\frac{\partial^2 P_{pert}(x,z,t)}{\partial t} \qquad (19)$$

Notice that the first three terms of each of the two summations equals zero. The expression then simplifies to

$$\sum_i \left[\left\{-2j\beta_i \frac{dA_i^+}{dz} + \frac{d^2 A_i^+}{dz^2}\right\} u_i(x) \exp[-j\beta_i z]\right] + c.c + \sum_i \left[\left\{-2j\beta_i \frac{dA_i^-}{dz} + \frac{d^2 A_i^-}{dz^2}\right\} u_i(x) \exp[j\beta_i z]\right] + c.c$$

$$= 2 \exp[-j\omega t] \cdot \mu \cdot \frac{\partial^2 P_{pert}(x,z,t)}{\partial t}$$

(20)

We also assume slow variations of the amplitudes

$$\left|\frac{d^2 A_i}{dz^2}\right| \ll \beta_i \left|\frac{dA_i}{dz}\right|$$

(21)

So, we can write,

$$\exp[j\omega t] \frac{1}{2} \sum_i \left[\left\{-2j\beta_i \frac{dA_i^+}{dz}\right\} u_i(x) \exp[-j\beta_i z]\right] + c.c$$

$$+ \exp[j\omega t] \frac{1}{2} \sum_i \left[\left\{-2j\beta_i \frac{dA_i^-}{dz}\right\} u_i(x) \exp[j\beta_i z]\right] + c.c = \mu \cdot \frac{\partial^2 P_{pert}(x,z,t)}{\partial t}$$

(22)

We multiply this equation with $u_i(x)$, and integrate over the cross section of the guide, and use mode orthogonality to arrive at our final result:

$$-\frac{dA_i^+}{dz} \cdot \exp[j(\omega \cdot t - \beta_i \cdot z)] + \frac{dA_i^-}{dz} \cdot \exp[j(\omega \cdot t + \beta_i \cdot z)] + c.c$$

$$= \frac{-j}{2\omega} \frac{\partial^2}{\partial t^2} \int_{-\infty}^{\infty} P_{pert}(x,z,t) \cdot u_i(x) dx$$

(23)

This equation can be used to treat a variety of waveguide structures with different types of interactions or coupling between guided modes. The exact form of the perturbation will depend on the waveguide structure at hand, but the general form of the coupled-mode equations will be the same.

## 2.4. Periodic Waveguides-Bragg Filters

We will now apply coupled mode theory to counter-propagating waves in a single mode waveguide with a periodic corrugation as shown in Fig. 4.

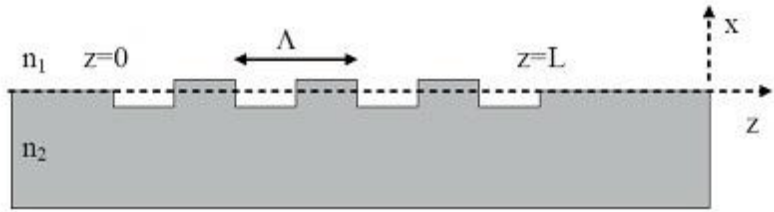

**Figure 4** Waveguide with periodic corrugation in one of the core-cladding interfaces. The waveguide is single mode, and we make the assumption that the only significant coupling is between counter propagating guided modes

The corrugation is scalar and we don't expect coupling between TE and TM modes, so in the following treatment we'll consider TE modes. We start by describing the field in the corrugated structure as a sum of the forward and backward propagating modes

$$E_y = A(z)u(x)\exp[j(\omega.t - \beta.z)] + B(z)u(x)\exp[j(\omega.t + \beta.z)] \quad (24)$$

where A and B are the amplitudes of the forward and backward propagating waves, and u(x) is the mode profile.

The perturbation in the corrugated region is

$$\vec{P}_{pert} = \Delta n(x,z)^2 \varepsilon_0 \vec{E} \quad (25)$$

We substitute the expression for the field into this expression to get

$$P_{pert} = \left(\frac{1}{2}\right)\Delta n(x)^2 \varepsilon_0 \{A(z)u(x)\exp[j(\omega.t - \beta.z)] + B(z)u(x)\exp[j(\omega.t + \beta.z)]\} \quad (26)$$

$$P_{pert} = \left(\frac{1}{2}\right)\Delta n(x)^2 \varepsilon_0 \, e^{j\omega.t} e^{-j\beta.z}\{A + Be^{-j2\beta.z}\}u(x) \quad (27)$$

Recall the fundamental coupled mode equation

$$-\frac{dA_i^+}{dz}.\exp[j(\omega.t - \beta_i.z)] + \frac{dA_i^-}{dz}.\exp[j(\omega.t + \beta_i.z)] + c.c$$

$$= \frac{-j}{2\omega}\frac{\partial^2}{\partial t^2}\int_{-\infty}^{\infty} P_{pert}(x,z,t).u_i(x)dx \quad (28)$$

which simplifies to

$$-\frac{dA}{dz} + \frac{dB}{dz}e^{j2\beta.z} = \frac{-j\omega.\varepsilon_0}{4}\{A + Be^{j2\beta.z}\}\int_{-\infty}^{\infty}\Delta n^2 u^2(x)\,dx \quad (29)$$

We will now assume that the corrugation has a square-wave shape as indicated in Fig. 4. The general conclusions are not dependent on the exact shape, so the following treatment, with appropriate adjustments, is valid also for non-square corrugations. The square-wave corrugations can be expressed as a series in the following form

$$\Delta n^2(x,z) = \Delta n^2 \sum_m C_m e^{j\cdot\frac{2m\pi.z}{\Lambda}} \quad (30)$$

$$C_m = \begin{cases} -\frac{j}{m\pi} & m = odd \\ 0 & m = even \end{cases} \quad (31)$$

By comparing this expression to the above coupled mode equations, we realize that only modes that are close to phase matched will experience significant coupling. In other words, we need only keep terms of the same periodicity. In a range of wave vectors, the equations can be simplified to

$$\frac{dA}{dz} = \frac{j\omega.\varepsilon_0}{4} Be^{j2\beta.z}C_m e^{-j\cdot\frac{2m\pi.z}{\Lambda}}\int_{-\infty}^{\infty}\Delta n^2 u^2(x)\,dx \quad (32)$$

$$\frac{dB}{dz} = \frac{j\omega.\varepsilon_0}{4} Ae^{-j2\beta.z}C_m e^{j\cdot\frac{2m\pi.z}{\Lambda}}\int_{-\infty}^{\infty}\Delta n^2 u^2(x)\,dx \quad (33)$$

$$\frac{dA}{dz} = K^* B e^{j2\Delta\beta.z} \quad (34)$$

$$\frac{dB}{dz} = KAe^{-j2\Delta\beta.z} \tag{35}$$

$$K = \frac{j\omega.\varepsilon_0}{4} C_m \int_{-\infty}^{\infty} \Delta n^2 u^2(x)\, dx \tag{36}$$

where

$$\Delta\beta = \beta - \frac{m\pi}{\Lambda} \tag{37}$$

Let us check energy conservation in the systems of equations we have found for modes in a Bragg grating. We start by deriving expression for the energies in the forward and backward propagating waves. Based on Eqns. 23 and 24 we can write

$$\frac{d}{dz}|A|^2 = \frac{d}{dx}(A.A^*) = A.\frac{dA^*}{dx} + A^*.\frac{dA}{dx} = A.B^*.Ke^{-j2\Delta\beta.z} + A^*.B.K^*e^{j2\Delta\beta.z} \tag{38}$$

$$\frac{d}{dz}|B|^2 = \frac{d}{dx}(B.B^*) = B.\frac{dB^*}{dx} + B^*.\frac{dB}{dx} = B^*.A.Ke^{-j2\Delta\beta.z} + B.A^*.K^*e^{j2\Delta\beta.z} \tag{39}$$

The difference between the rates of change in the forward-propagating and backward-propagating energy is then

$$\frac{d}{dz}|A|^2 - \frac{d}{dz}|B|^2 = A.B^*.Ke^{-j2\Delta\beta.z} + A^*.B.K^*e^{j2\Delta\beta.z}$$
$$- B^*.A.Ke^{-j2\Delta\beta.z} + B.A^*.K^*e^{j2\Delta\beta.z} = 0 \tag{40}$$

We see that the rate of change in forward-propagating energy is exactly balanced by the rate of change in backward-propagating energy, which is the correct result for loss-less, counter-propagating waves.

## 2.5. Modes of the Bragg Grating

The set of equations describing the modes of the Bragg Grating (Eqns. 32-36) can now be solved. Assuming that the forward propagating mode has an amplitude $A_0$ at z=0, and that the backward propagating wave is zero at z=L, we find

$$A = A_0 e^{j\Delta\beta.z} . \frac{-\Delta\beta \sinh[S(z-L)] + j S\cosh[S(z-L)]}{-\Delta\beta \sinh[SL] + j S\cosh[SL]} \tag{41}$$

$$A = A_0 e^{j\Delta\beta.z} . \left[\frac{-\Delta\beta \sinh[S(z-L)]}{-\Delta\beta \sinh[SL] + j S\cosh[SL]} + \frac{j S\cosh[S(z-L)]}{-\Delta\beta \sinh[SL] + j S\cosh[SL]}\right] \tag{42}$$

$$B = A_0 . jK . e^{-j\Delta\beta.z} . \frac{\sinh[S(z-L)]}{-\Delta\beta \sinh[SL] + j S\cosh[SL]} \tag{43}$$

where

$$S = \sqrt{K^2 - \Delta\beta^2} \tag{44}$$

When $\Delta\beta = 0$, this simplifies to

$$A = A_0 . \frac{\cosh[K(z-L)]}{\cosh[KL]} \tag{45}$$

$$B = A_0 . \frac{\sinh[K(z-L)]}{\cosh[KL]} \tag{46}$$

## 3. Results and Discussion

With $Al_xGa_{1-x}N$/GaN material composition considered as a unit block of the periodic organization $\Lambda$, we have succeeded in fabricating the $Al_xGa_{1-x}N$ layer with a quantum width of 244.4nm. The quantum well is then filled with GaN material deposition. However, we should note that here we are making the implicit assumption that the lengths of the high and low index regions ($L_H$ and $L_L$) are the same. From the practical fabrication point of view, we should note that the high index region should be shorter ($L_H<L_L$), but if we insist on matching both physical ($\Lambda=L_H+L_L$) and optical lengths ($n\cdot\Lambda= (n+\Delta n)L_H+(n-\Delta n)L_L$), then it follows that the two regions must have the same length ($L_H = L_L$). This is solely for the comfort of our calculation.

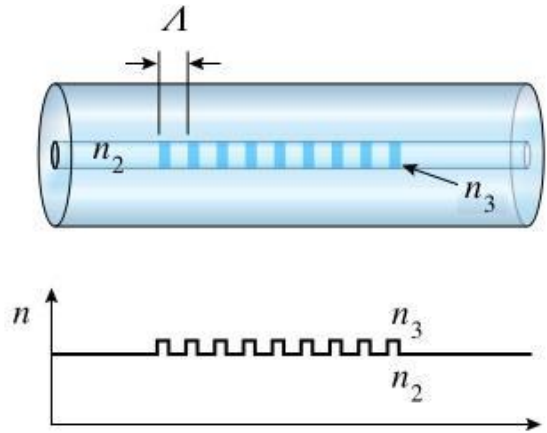

**Figure 5** A Fiber Bragg Grating structure with refractive index profile

However, Fig. 6 and Fig.7 shows us a corrugated structure because in reality, $L_H = L_L$ is not possible to achieve.

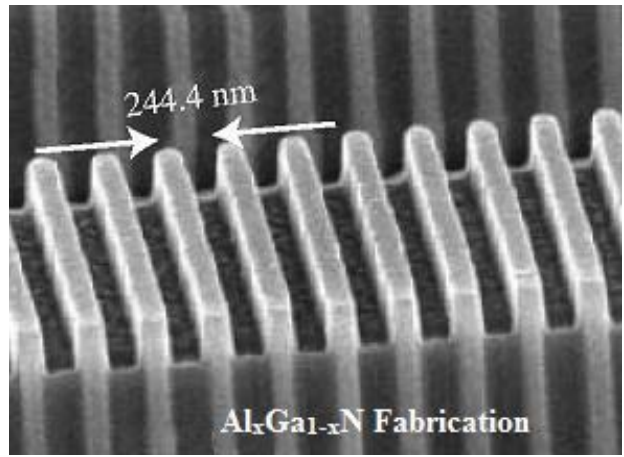

**Figure 6** Microscopic view of $Al_xGa_{1-x}N$ fabrication

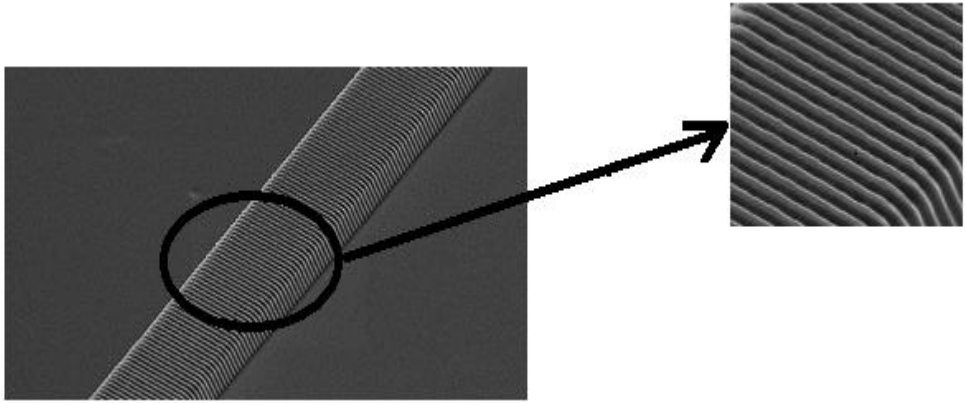

**Figure 7** Microscopic View of the Bragg grating. A specific part of the structure is highlighted to emphasis on the fact that $L_H \neq L_L$ forming a corrugated structure.

The forward wave modes are plotted as a function of grating wavelength for a coupling coefficient of K=0.1cm$^{-1}$ as shown in Fig.8 and Fig.9,

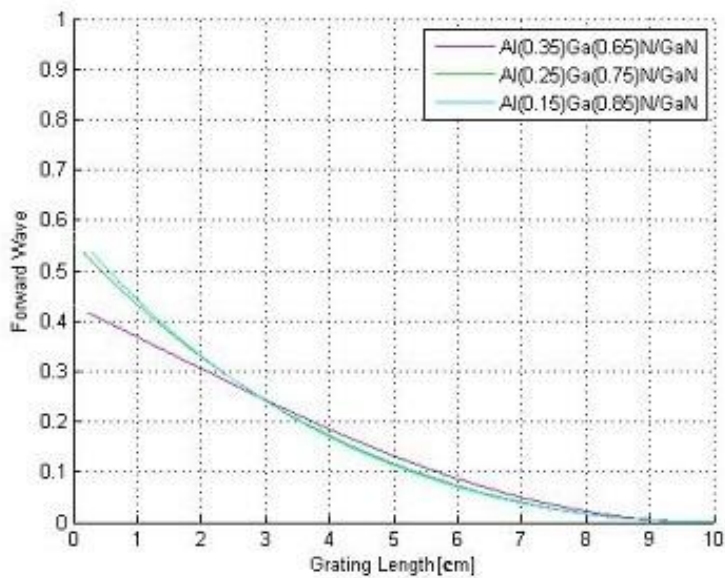

**Figure 8** K=0.1cm$^{-1}$, Grating Length=10 cm

With a total grating length of 10cm, the forward wave decays almost linearly when index difference is higher, i.e., for high Al composition whereas the corresponding wave decreases in an exponential manner when the grating length is increased to twice the original length. Another result which is of interest here, is that, composition for Al should range from x=0.1 to x=0.3 otherwise we see anomalous behavior in the magnitude of the forward wave for x=0.35 when the grating length is increased. The plots adhere to the expression we, got for the power in the forward wave given by Eqn. (42). The mathematical and the physical significance of the expression have already been described in the preceding section.

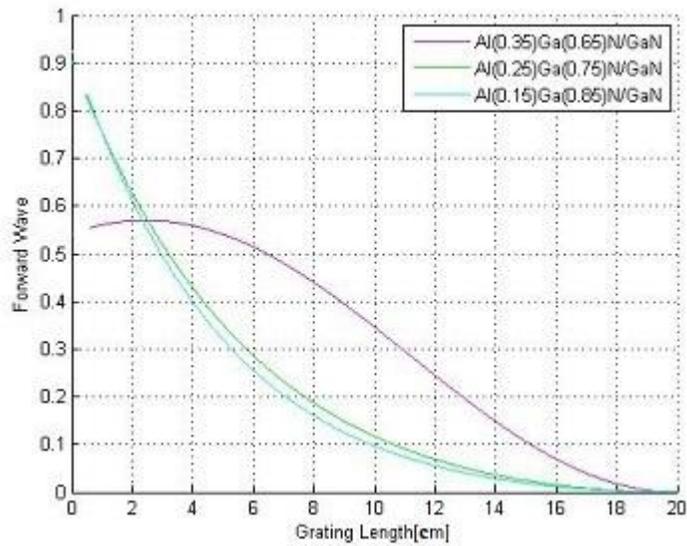

**Figure 9** K=0.1cm$^{-1}$, Grating Length=20 cm

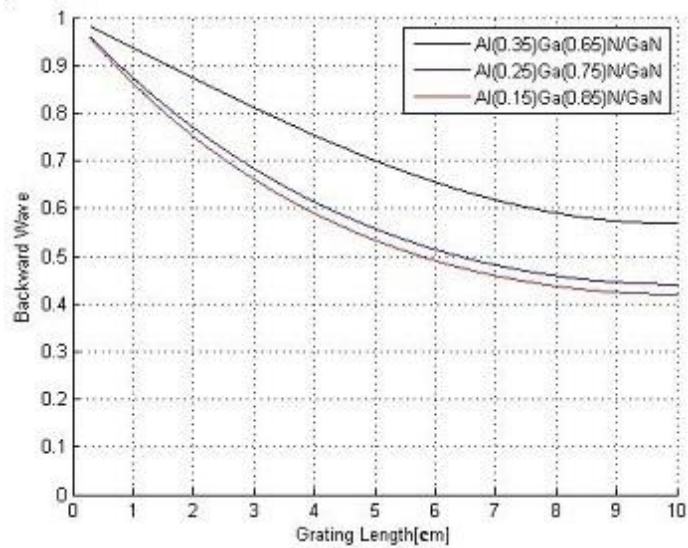

**Figure 10** K=0.1cm$^{-1}$, Grating Length=10 cm

Looking at the plots shown in Fig. 10 and Fig. 11 backward wave profiles of the optical waveguide structure, with a coupling coefficient of K=0.1cm$^{-1}$ (similar to the previous case), it is noticeable that for a coupling coefficient of K=0.1cm$^{-1}$, backward wave remains higher compared to the forward propagating waves. Also for higher mole fraction (x), wave increases slightly with increasing grating length, and then starts to decrease very slowly. This is because weak coupling between propagating waves transfers energy from forward wave to backward wave, and thus its energy increases.

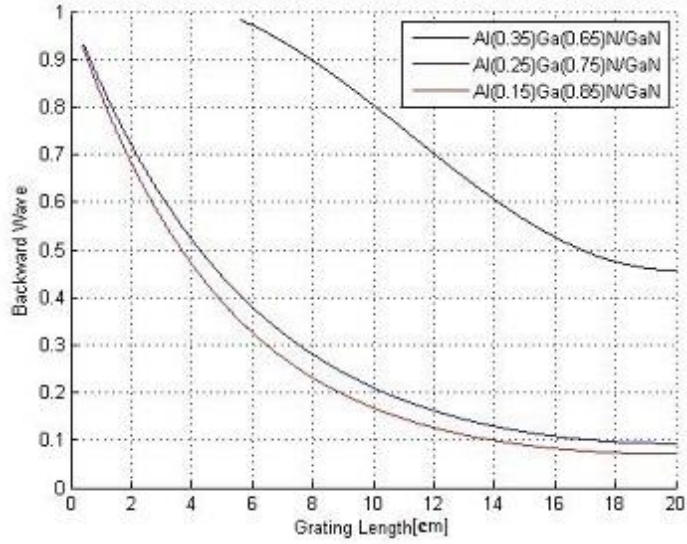

**Figure 11** K=0.1cm$^{-1}$ , Grating Length=20 cm

However, there is a similarity in the characteristic curve between both of the forward and the backward wave i.e. with increase in the grating length; the magnitude falls exponentially, in accordance with Eqn. (39).

Coming to both Eqn (41) & Eqn (43), restating both of them:

$$A = A_0 e^{j\Delta\beta z} \cdot \frac{-\Delta\beta \sinh[S(z-L)] + j S\cosh[S(z-L)]}{-\Delta\beta \sinh[SL] + j S\cosh[SL]} \quad (41)$$

$$B = A_0 \cdot jK \cdot e^{-j\Delta\beta z} \cdot \frac{\sinh[S(z-L)]}{-\Delta\beta \sinh[SL] + j S\cosh[SL]} \quad (43)$$

Eqn (41) can be rewritten as

$$A = A_0 e^{j\Delta\beta z} \cdot \left[\frac{-\Delta\beta \sinh[S(z-L)]}{-\Delta\beta \sinh[SL] + j S\cosh[SL]} + \frac{j S\cosh[S(z-L)]}{-\Delta\beta \sinh[SL] + j S\cosh[SL]}\right] \quad (42)$$

Simple mathematics concept tells us that near similar expressions of a function relate to a somewhat similar nature of their corresponding curves. Since we have plotted the magnitudes as a function of grating length, the magnitudes of both |A| and |B| are similar. However, higher value of |B| results due to an extra term i.e. the complex term

$$\frac{j S\cosh[S(z-L)]}{-\Delta\beta \sinh[SL] + j S\cosh[SL]}$$

which adds some significant component to the magnitude of the backward wave. Moreover multiplication of K (coupling coefficient) with the expression in eqn. (40) attenuates the forward wave since 0<K<1.

Having discussed the following points, we now seek to study both the forward and backward wave profiles on a single plot, for a better understanding of the reflection and transmission in optical waveguide structure, with varying grating length.

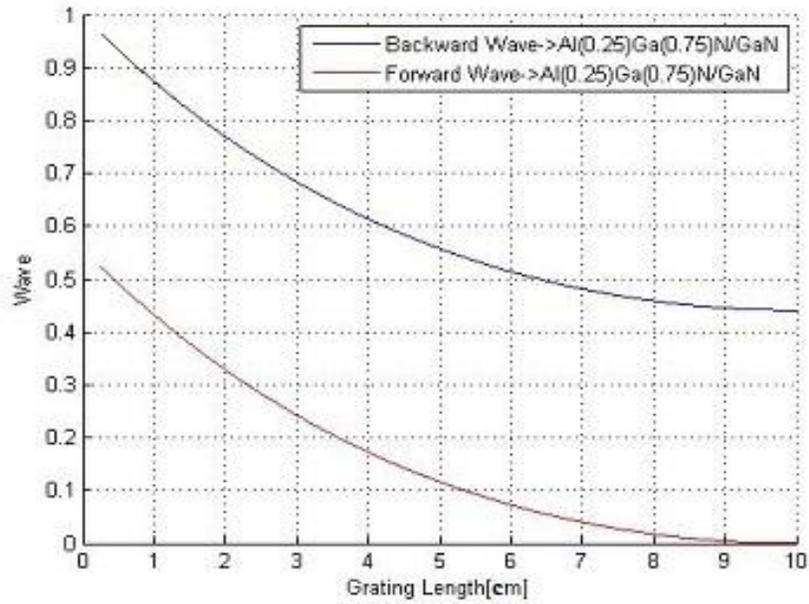

**Figure 12** K=0.cm$^{-1}$ , L=10 cm

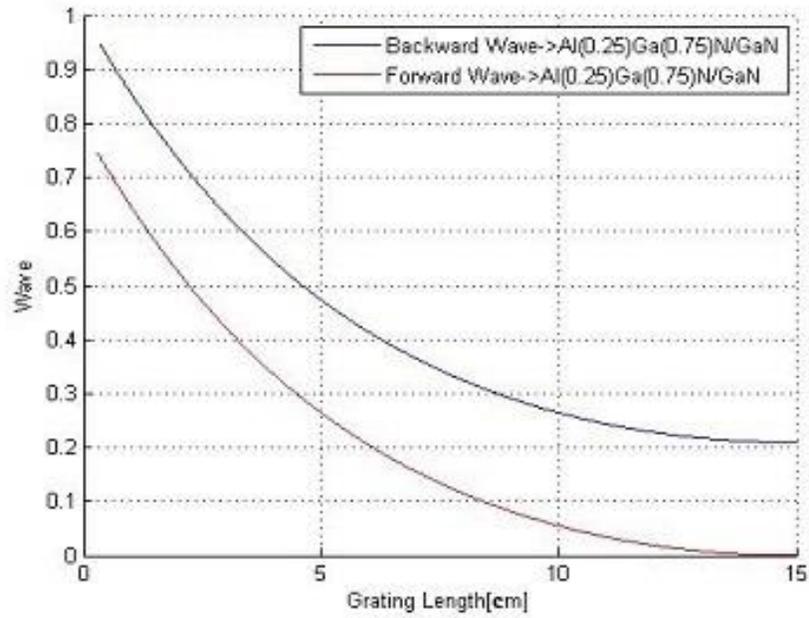

**Figure 13** K=0.cm$^{-1}$ , L=15 cm

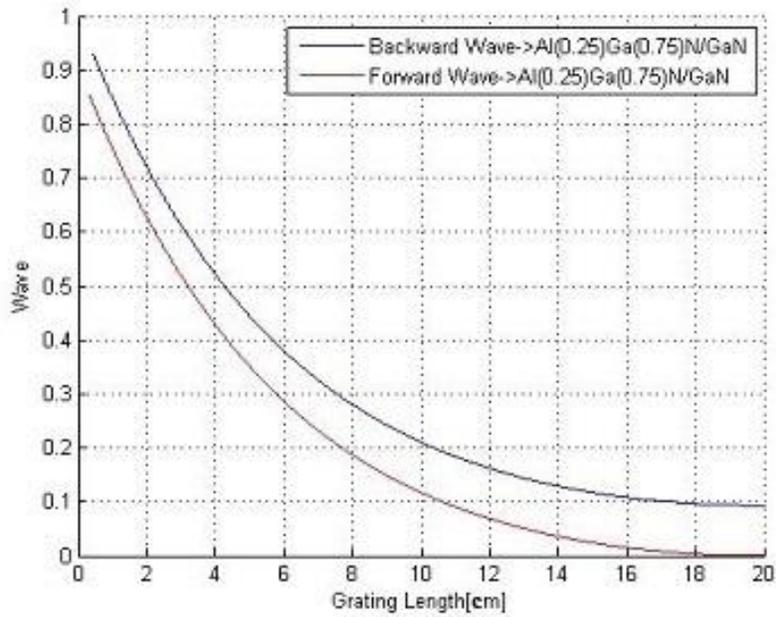

**Figure 14** K=0.cm$^{-1}$ , L=20 cm

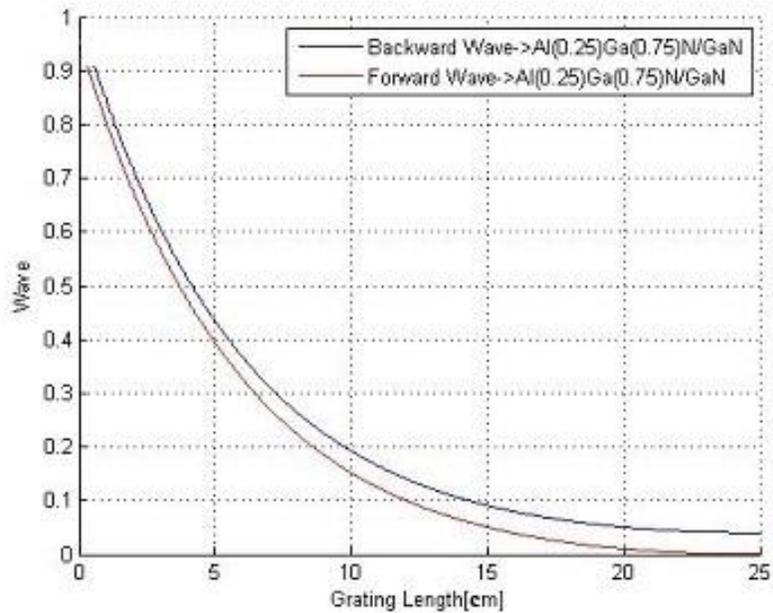

**Figure 15** K=0.cm$^{-1}$ , L=25 cm

The graphs in Fig.12-15 plot the relation between the forward and backward wave profiles with varying grating length.

Here $Al_{0.25}Ga_{0.75}N/GaN$ was taken as a single block of the composition (the most suitable fabrication values for mole fraction, x=0.25) after studying the various forward and backward propagating modes with different material composition. As the corrugated region gets longer, more of the power is coupled into the reflected wave. This is what we would expect; a longer grating leads to smaller transmission and larger reflection, which means that the forward and backward propagating fields are closer in magnitude. In the extreme case of an infinite grating, the forward and backward propagating waves are equal at all points in the grating.

## 4. Conclusion

In this paper, we study fiber-optic devices that are important in Optical MEMS and Nanophotonics. The first class of devices that is described simply provides a means for coupling to optical fibers and waveguides. In summary, wave propagating modes in an optical periodic waveguide structure namely the fiber Bragg filter is analytically computed for different grating length for a fixed coupling coefficient.

It is observed that as the grating length increases, the magnitude of the reflected wave increases as more power is coupled into. This is what we would expect; a longer grating leading to smaller transmission and larger reflection, which means that the forward and backward propagating fields are closer in magnitude. Also with increase of Al mole fraction (x), difference of refractive index increases, so magnitude of becomes higher for backward and lower for forward waves respectively.

These characteristic curves can be utilized to study how waves propagate through the optical waveguides which have a special place in optical communications for several reasons: (1) They are used in their elementary form in many optical systems, (2) they form the basis for a large number of advanced optical devices, and (3) their descriptions build intuitive understanding of central optics concepts, including photonic bandgap structures.